\documentclass[twocolumn,twoside,slac_two]{revtex4}
\usepackage{graphicx}
\usepackage{fancyhdr}
\begin{document}

\title{Fermi-LAT observations of GRBs with weak LAT emission}

%

\author{Masanori Ohno}
\affiliation{ISAS/JAXA, 3-1-1 Yoshinodai, Sagamihara, Kanagawa, 229-8510, Japan}
\author{on behalf of the Fermi-LAT/GBM collaborations}

\begin{abstract}
We present the analysis results of three Gamma-Ray Bursts (GRBs) detected by the
Gamma-ray Burst Monitor (GBM) and the Large Area Telescope (LAT) onboard Fermi:
the two long GRB 080825C and GRB 090217, and the first short burst with GeV photons GRB 081024B.
The emission from GRB 081024B observed by the LAT above 100 MeV is delayed with respect to the
GBM trigger, and significantly extends after the low-energy episode. Some hints for spectral
hardening was observed in this burst as well as in GRB 080825C, possibly related to a separate
and harder component showing up at late times. Conversely, GRB 090217 does not exhibit any
noticeable feature. Together with the other bright LAT detected bursts
(e.g. GRB 080916C and GRB 090510), these observations help to classify the GRB properties
and give new insight on the acceleration mechanisms responsible for their emission at the
highest energies.

\end{abstract}

\maketitle

\thispagestyle{fancy}


\section{Introduction}

Recent observational progresses revealed that
Gamma-ray Bursts (GRBs) are extremely energetic explosion
originated by core collapse of the massive stars or
merger of binary of neutron stars or black holes at
cosmological distance. The gamma-ray prompt emission below
several MeV band has been usually observed but the detection
of high-energy emission above 100 MeV was reported only a
few times by the Energetic Gamma-Ray Experiment Telescope (EGRET)\cite{Dingus}
and recently by Astro-rivelatore Gamma a Immagini LEggero (AGILE)\cite{Giuliani}.
The observed properties of these high-energy photons from some GRBs
showed distinct spectral and temporal behavior compared with low
energy photons below several MeV, suggesting different gamma-ray
emission processes between low and high-energy photons\cite{Hurley}\cite{Gonzaretz}.
In addition, comparing observed properties of high-energy photons
between short and long duration GRBs will help to study the sub
classes of GRBs. However, very low sample number of
GRBs with high-energy photons before Fermi era makes it difficult to
reveal the detailed properties of high-energy emission from GRBs.

\begin{figure}
\rotatebox{-90}{\includegraphics[width=40mm]{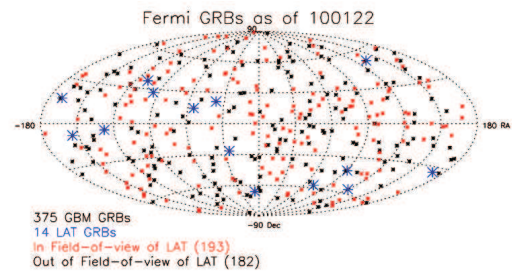}}
\caption{GRB skymap detected by Fermi between 14th July, 2008 to 22th January, 2010. GRBs detected by
  the GBM in field of view of the LAT and out of
field of view of the LAT are shown by red and black asterisks, respectively.
GRBs detected $>$100 MeV photons by the LAT are
shown by large blue asterisks}
\label{grb_skymap}
\end{figure}

The Fermi Gamma-ray Space Telescope has successfully detected 14 GRBs
with high-energy photon ($>$100 MeV) so far thanks to its unprecedented effective
area above 100 MeV of the Large Area Telescope (LAT)\cite{Atwood}.
Figure\ref{grb_skymap} shows the
sky distribution of all GRBs detected by the LAT and Gamma-ray Burst Monitor (GBM)
as of 22th January, 2010.
Fermi firmly confirmed the distinct high-energy spectral component with
respect to traditional Band function in low energy band observed by the
GBM from some bright GRBs so far (GRB 090510 and  GRB 090902B). Furthermore, high-energy photon
detected by short-hard GRB 090510 provides a stringent photon dispersion limit,
which strongly disfavors the quantum gravity model of space-time causes a linear
variation of the speed of light with photon energy\cite{grb090510}.
In addition to those interesting properties for individual LAT GRBs, the large
number of samples enable us to study the systematic properties of 
high-energy emission from GRBs. Actually, the high-energy photons of the LAT
are often delayed and extended compared with the low-energy emission from GRBs,
suggesting a different acceleration process and/or emission process in high
energy emission. And also Fermi has detected high-energy photons from both
short and long duration GRBs. This would help to classify the sub classes of
GRBs with high-energy emission properties.
In this paper, we present the detail analysis result of three GRBs with
weak LAT detection, GRB 080825C\cite{grb080825c}, GRB 081024B\cite{grb081024b},
and GRB 090217\cite{grb090217} and discuss
the global properties of the high-energy emission of GRBs by comparing with
other LAT bright GRBs.

\section{Analysis Techniques}

We have performed a detailed temporal and spectral analysis for three GRBs with weak
LAT emission through the careful procedures, which were developed by our collaborations
for event selection, selection of the region of interest (ROI), and background
estimation. In this section, we briefly summarize these
important procedures for our analysis. See Abdo et al. 2009a for more detail.

\subsection{Event Selection}

The standard selection of the LAT photon event has been developed by Atwood et al. 2009.
There are three event classes, named as ``diffuse'', ``source'', and ``transient''
class for specific scientific analysis. In the case of GRB observations, the smaller region of the
sky and shorter time scale compared with diffuse sources allows the event section to be
relaxed, and the ``transient'' event class is usually used to detection, localization, and
analysis for the prompt emission. On the other hand, diffuse class, which is used to analysis
a faint source with longer time interval and covering larger region of the sky
is also used for the afterglow search of GRBs. In addition to these standard event classes,
we developed more optimized event selection than the ``transient'' class for the spectral
analysis based on Monte-Carlo simulations. From this study, we found that more relaxed can be
used because the background contamination is less issue due to short time windows of GRB analysis.
This optimized event class, so-called ``S3'' class is used for the spectral analysis of
GRB 080825C. For other GRBs, GRB 081024B and 090217, the standard ``transient'' class is applied
for the spectral analysis.

\subsection{Energy-dependent ROI}

The LAT point-spread function (PSF) strongly depends on the incident energy and the conversion
point on the tracker, and thus the detector response of the LAT is separated into ``FRONT'' and
``BACK'' events. We also consider the energy-dependent region of interested (ROI) for these ``FRONT''
and ``BACK'' event individually based on the 95 \% containment radius (PSF95) and the 95 \% LAT
localization error (Err95) as follows:
\begin{equation}
  ROI(E) = \sqrt{PSF95(E)^2+Err95^2}
  \end{equation}
The maximum value of this energy-dependent ROI is set to 10 and 12 degree for ``FRONT'' and ``BACK'' event, respectively not to contain too large background region. We apply this energy-dependent ROI
for all three GRBs.

\subsection{Background Estimation}

Because of very small number of photons detected for three GRBs presented here, the
background estimation should be performed very cautiously. The background rate
strongly depends on many parameters such as the incident angle of the burst or
the position in the instrument.
Therefore, it is not straightforward to estimate the accurate LAT background using
off-source region around the trigger time.
 There are two components of LAT background events;
cosmic-ray (CR) background and diffuse gamma-ray background (from extra-galactic and Galactic).
Since these components have different properties, we have developed two different methods to estimate
the expected background rate. The amount of gamma-ray background only depends on the exposure
of a certain direction in the celestial sphere. Therefore, this component can be estimated
by a simple scaling of the number of gamma-rays detected in longer observations. Here, we
used six months of LAT data for this scaling. The amount of CR background also depends on
the exposure of specific direction of the sky, however, this component also changes with the
geomagnetic coordinates at the location of the spacecraft, and this dependence cannot be
estimated by the same way as the gamma-ray background. Thus, we utilized Monte Carlo simulation
of the GRB observation to estimate the CR background.

\section{Analysis Results of Three LAT GRBs}

\subsection{Detection and Localization}

GRB 080825C is the first long GRB detected by the LAT\cite{Bouvier}.
The GBM triggered on this burst at 14:13:48 UT\cite{van der Horst}.
We estimated the burst location using ``transient'' class described above.
In this analysis, we fit the LAT data assuming the power-law shape of the
point source. To estimate the error of the position of this point source,
we calculate the test statistics (TS), assuming point source for each grid
of the map. As the result of this analysis, we found the best-fit position
of this GRB at (RA,Dec)=(233.9, -4.5) with
0.8 degree of 68 \% error radius. The detection significance is estimated by
several independent method such as unbinned likelihood analysis or some Bayesian
approaches, and we obtained about 6 sigma by these methods. Detailed description
of these methods can be found in Abdo et al. 2009a.

The short GRB 081024B triggered the GBM at 21:22:41 UT\cite{Connaughton} and the LAT ground
analysis found high-energy photons from this burst up to 3 GeV\cite{Omodei}. Thus, this 
is the first short GRBs which is detected GeV photon.
Through the same procedure as GRB 080825C using ``transient'' class events,
the LAT location of this GRB was estimated as  (RA,Dec)=(322.86, 21.16) with 0.22
degree of 68 \% error, and
the detection significance of this burst is 6.7 sigma. 

Another long GRB, 090217 triggered the GBM at 04:56:42 UT\cite{von Kienlin}. This burst firstly
detected by the blind search of the LAT data by the on-ground
Automated Science Processing (ASP), and this is confirmed by the follow-up analysis
around GBM position\cite{Ohno}. The ``transient'' class events are used to obtain the location and
significance of the LAT data same as GRB 080825C, and the best-fit position is found to
be (RA, Dec)=(204.74, -8.43) with 0.37 degree error radius. The detection significance is
 8.4 sigma and 9.2 sigma with the semi-Bayesian method and maximum TS value obtained by
unbinned likelihood fit, respectively.

 \subsection{Temporal Properties}
Figure \ref{latgbmlc} shows the energy-resolved light curve of three LAT GRBs.
The top two panels in each figure show the background-subtracted light curve
of NaI and BGO detectors of GBM. The LAT events around the GBM position
are plotted on the other bottom panels in each figure. 
From these light curves, we can see various temporal properties of high-energy photons
from bursts to bursts. First, we can clearly see that the onset of LAT high-energy photons
are delayed with respect to the GBM low energy photons for GRB 081024B. In the case of
GRB 080825C, this delay can be seen but it is not statistically significant.
Furthermore, the LAT high-energy photons last longer than GBM low energy emission;
$\sim$35 s for GRB 080825C, and $\sim$3 s for GRB 081024B. The highest energy photon, 572 MeV
for GRB 080825C and 3.1 GeV for GRB 081024B are detected when the low energy emission becomes
weak. Whereas the LAT high-energy photons shows various properties different from GBM
low energy photons for GRB 080825C and GRB 081024B, few noticeable temporal features can be
observed from GRB 090217: there is no delay and extended emission of high-energy photons.

\subsection{Spectral Properties}

\begin{figure}[!tbp]
\centering
\includegraphics[width=80mm]{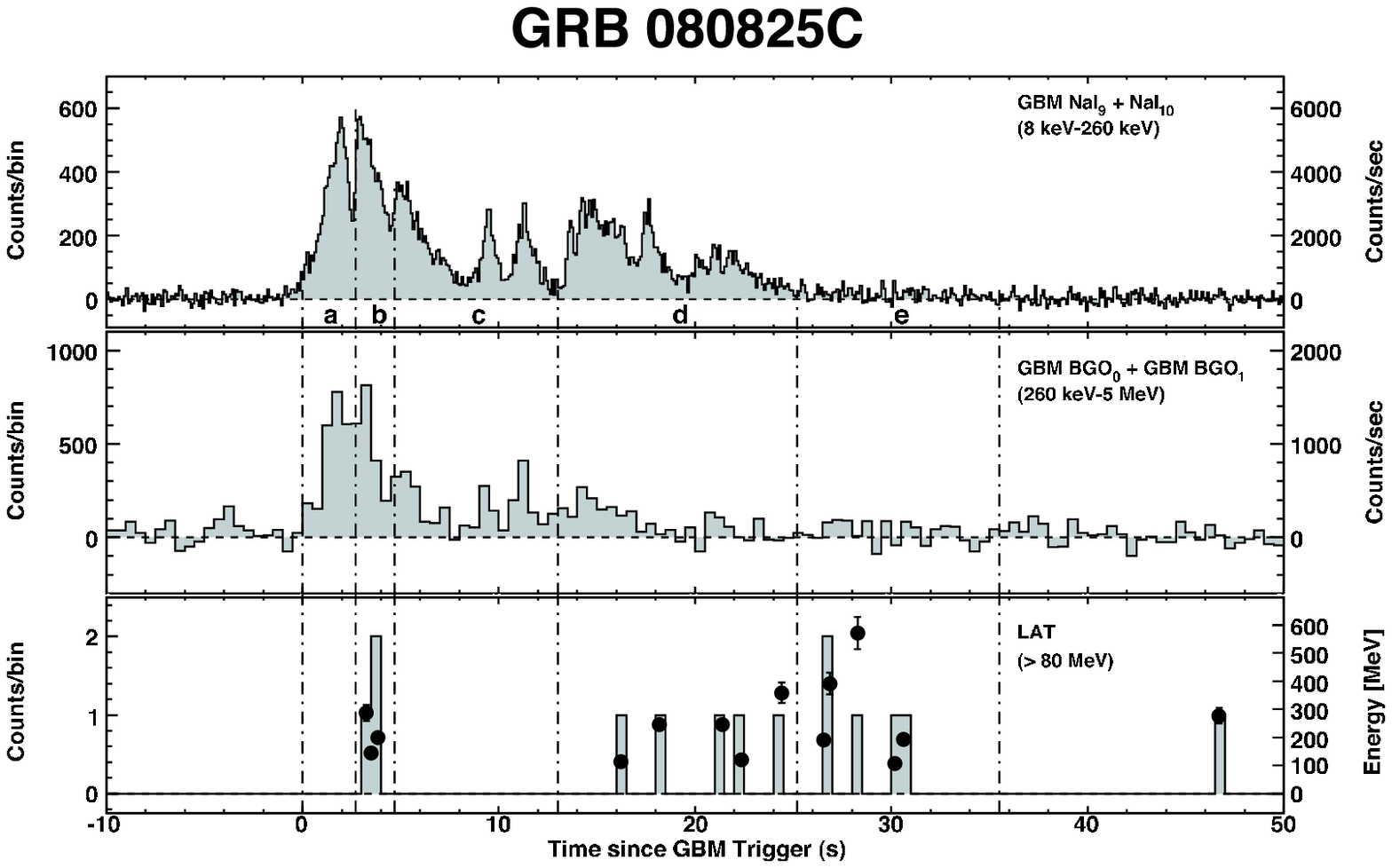}
\includegraphics[width=80mm]{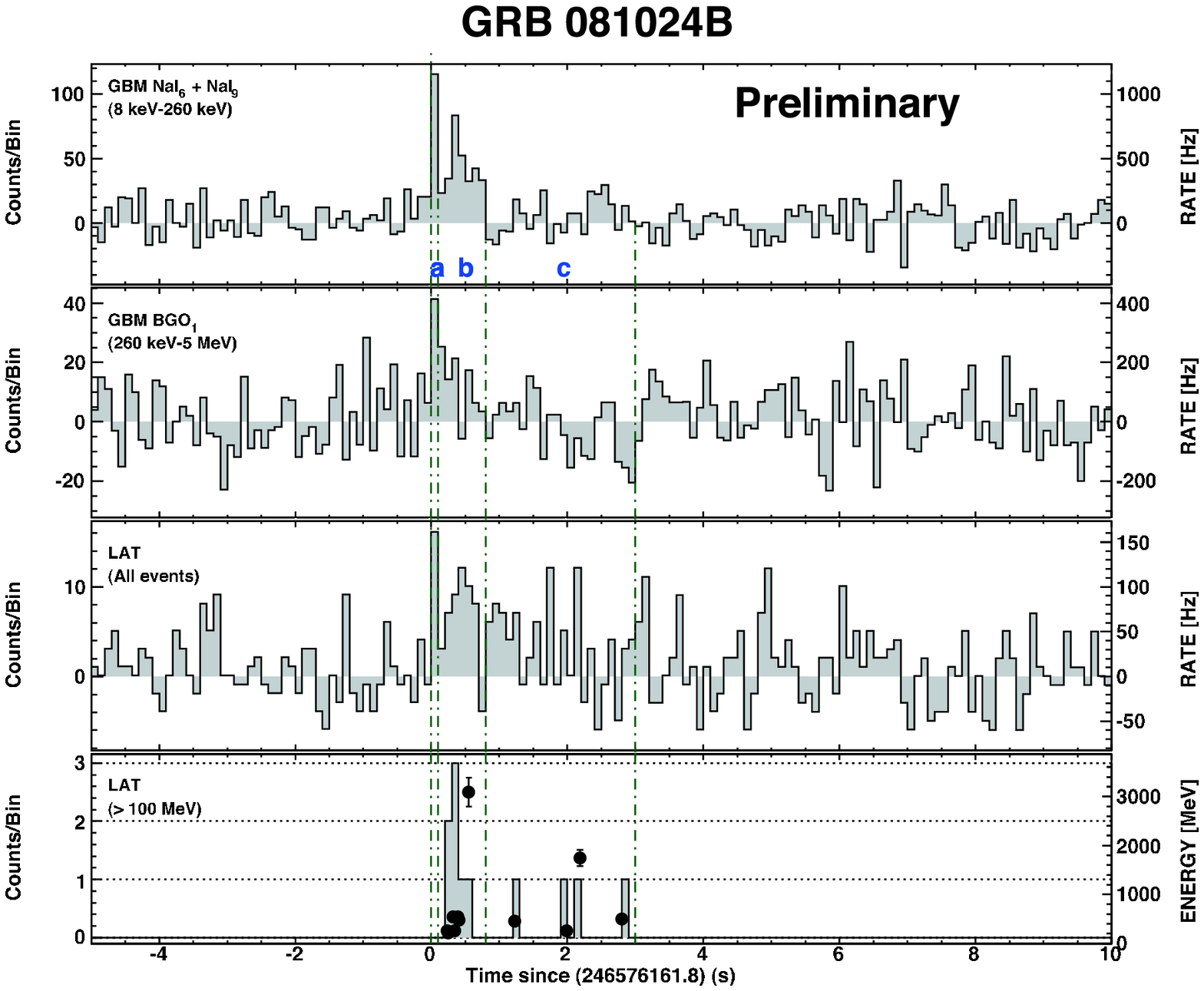}
\includegraphics[width=80mm]{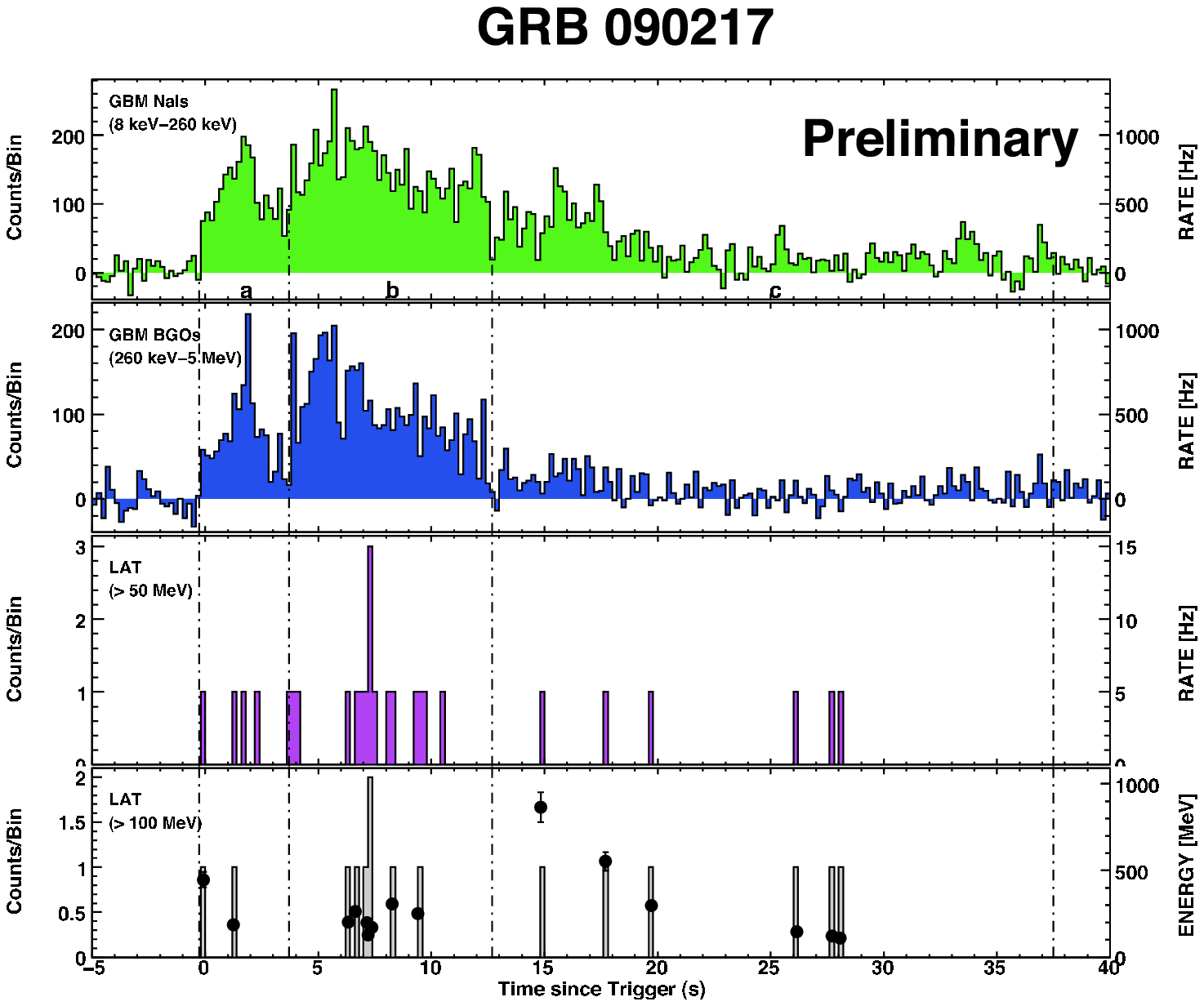}
\caption{Energy-resolved light curves of three weak LAT GRBs. Top two panels are background subtracted
GBM light curve and the other bottom panels are the LAT light curves. The LAT events are the ``S3'' events above 80 MeV for GRB 080825C,
all ``transient'' events and ``transient'' events above 100 MeV for GRB 081024B, and
``transient'' events above 50 MeV and above 100 MeV for GRB 090217, respectively. Black dots in each figure show the energy of each LAT event with 1 $\sigma$ error. The vertical dashed-line in each figure shows the time-interval used in the time-resolved spectral analysis.} 
\label{latgbmlc}
\end{figure}

\begin{figure}[!h]
\centering
\includegraphics[width=75mm]{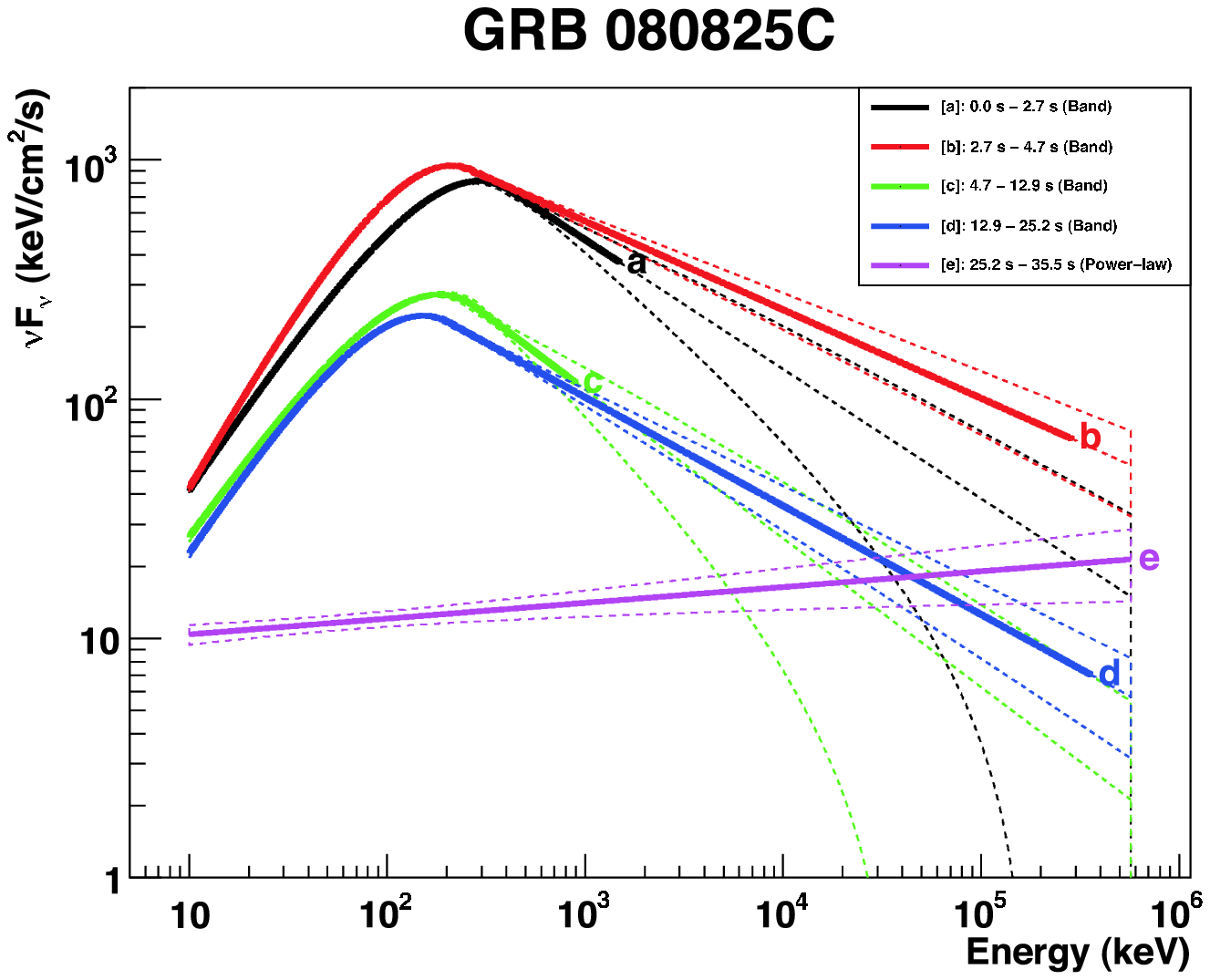}
\includegraphics[width=82mm]{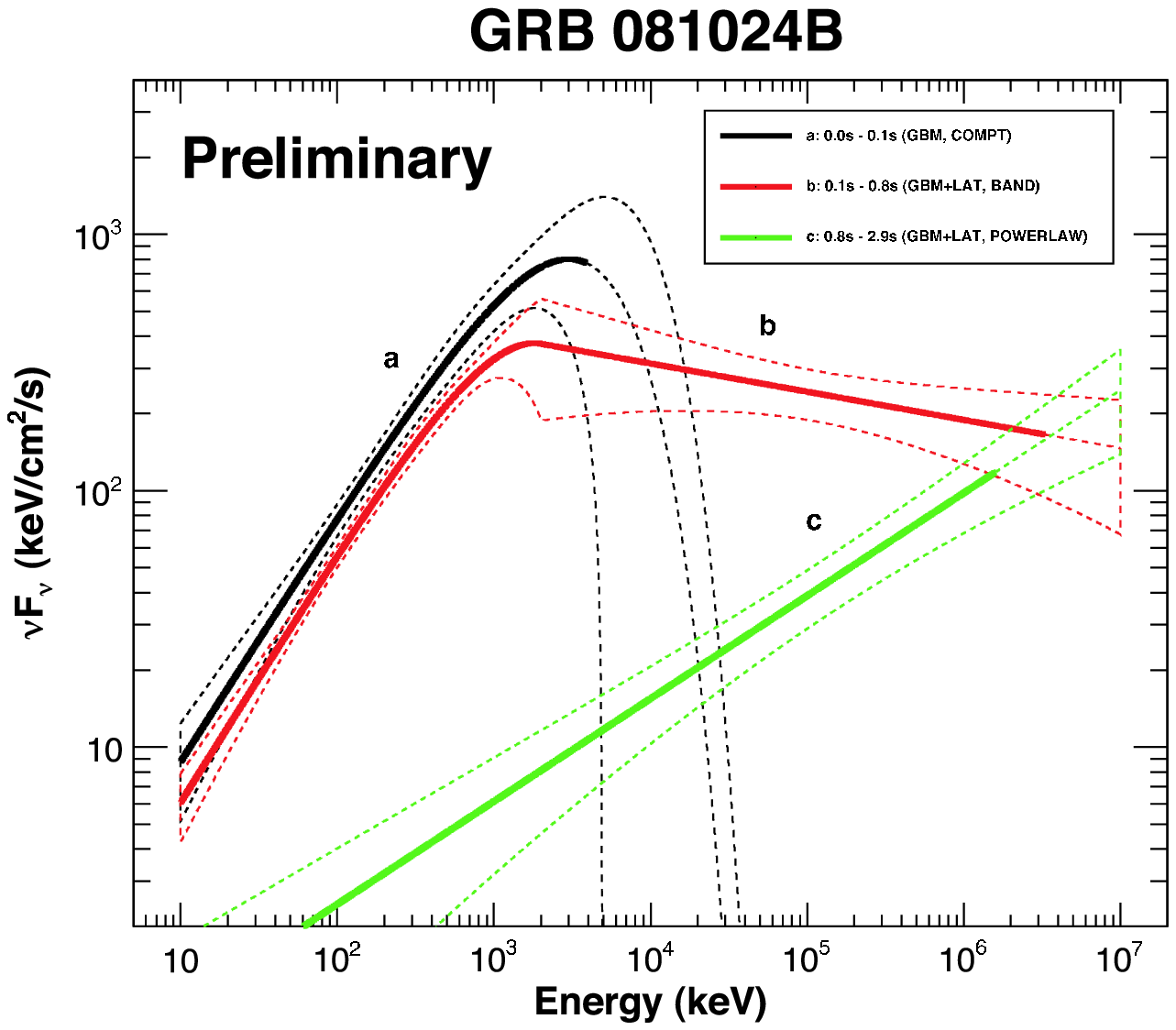}
\includegraphics[width=85mm]{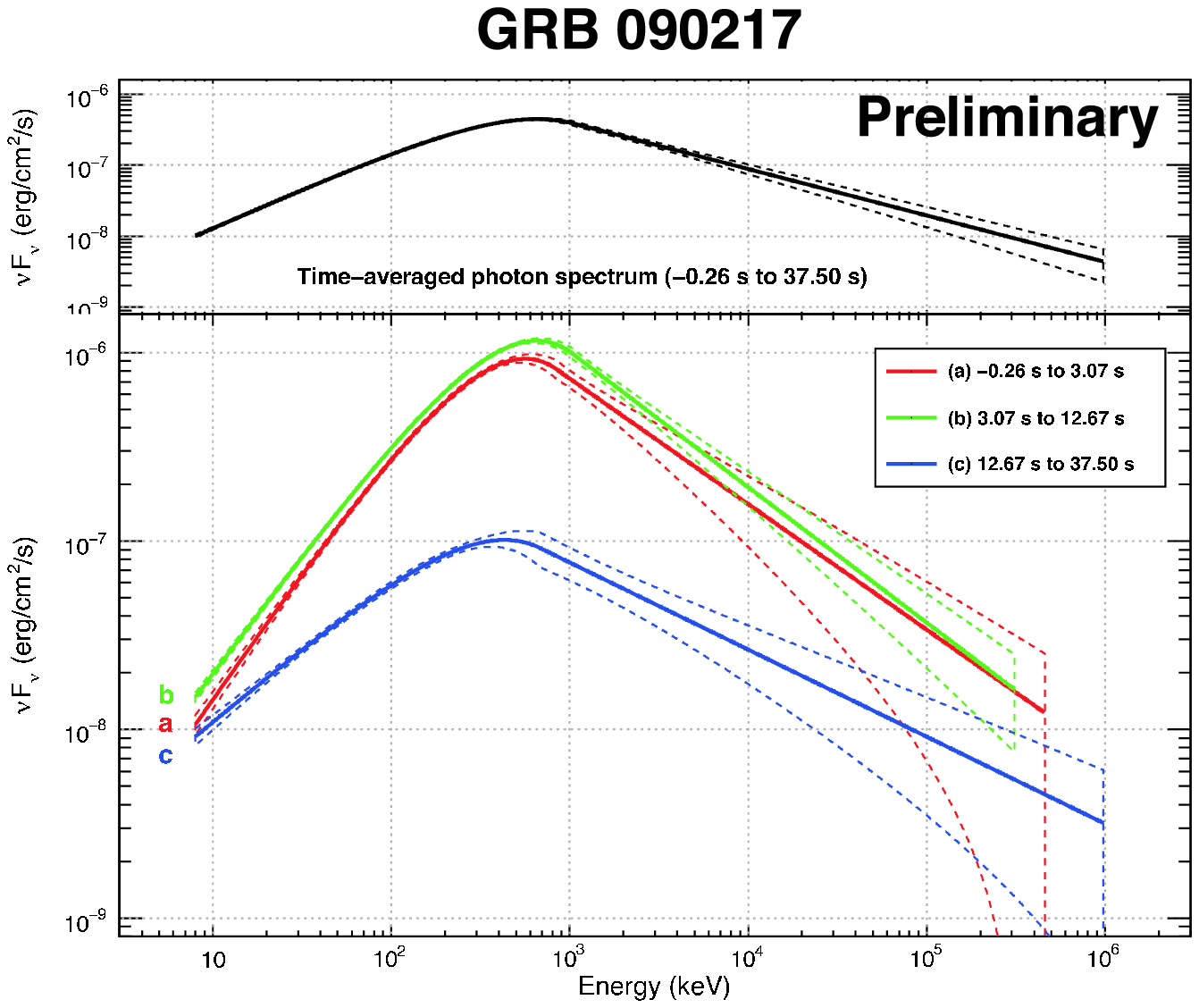}
\caption{The best-fit models obtained by the time-resolved spectral analysis for three weak LAT GRBs.
Solid lines in each figures show the best-fit model. The dashed-lines around these best-fit models represent the contour level at the 68 \% confidence interval. Each solid lines corresponds to time-resolved spectra which are extracted in the time interval shown in figure \ref{latgbmlc}.}
\label{latgbmspec}
\end{figure}

The time-resolved spectral analysis combining the LAT and GBM
data was performed for three weak LAT GRBs.
Figure \ref{latgbmspec} shows the resulting spectral models in the
$\nu$F$_\nu$ representation. Each figure shows the
best-fit model in each time interval shown in figure \ref{latgbmlc}.
From this analysis, we found no significant evidence of an additional
high-energy spectral component such as extra power-law component or
high-energy cut-off structure, and the single
Band function or Comptonized model give a good fit for both time-integrated
and time-resolved spectrum for all three weak LAT GRBs. We only found a weak
evidence of high-energy cut-off from GRB 080825C, E$_{\rm cut}=1.77^{+1.59}_{-0.56}$ MeV
at the first time bin with the significance level of 4.3 sigma. 
Interestingly, significant spectral hardening at the high-energy band in the last
time bin can be seen for GRB 080825C and GRB 081024B, where the GBM low energy
emission is almost back to background level. The spectrum of this late-time high
energy emission can be represented by a single power-law model with the photon index of
$-1.95\pm0.05$ and $-1.59^{+0.2}_{-0.07}$ for GRB 080825C and GRB 081024B, respectively.
This late-time high-energy component is common to other bright LAT GRBs and the possible
origin of this emission is synchrotron self-Compton emission during
afterglow phase or cascades induced by ultrarelativistic hadrons accelerated by the
relativistic jet. 
GRB 090217, however, shows no significant evidence of additional spectral component and
spectral evolution in any time interval unlike GRB 080825C and GRB 081024B.

\section{Comparison with other bright LAT GRBs}
Table \ref{grbcomp} shows the comparison of the various properties between weak and bright LAT
GRBs. Above temporal and spectral analysis for three weak LAT GRBs revealed that
the delayed onset and long-lived behavior of high-energy photons from
GRB 080825C and GRB 081024B. Such behavior is also reported from other bright
LAT GRBs such as GRB 080916C, 090510, and 090902B, and we might think that
the delayed onset and long-lived high-energy photon is the common feature of
GRBs. However, we can not find any temporal and spectral feature from GRB 090217.
Such ``featureless'' feature of GRB 090217 may suggest the unique mechanism of the
broad band gamma-ray emission or indicate that the different high-energy emission
class exist. When we compare the high-energy emission properties between short and
long duration GRBs, there is no clear difference between these two classes in both
weak and bright LAT GRBs.
More GRB observations are needed to investigate common and uncommon properties and
possible classification in high-energy emission of GRBs, and such study would be
important to give new insight on the acceleration mechanisms for the high-energy emission
of GRBs.

\clearpage

\begin{table*}[t]
\begin{center}
\caption{Comparison of properties of high-energy (HE) photons of weak LAT GRBs with other bright LAT GRBs.}
\begin{tabular}{|c|c|c|c|c|c|c|c|c|}
\hline
GRB & short/long & num of events  & num of events& delayed & long-lived & extra comp. & highest energy & redshift \\
    &  class        & $>$ 100 MeV   &  $>$ 1 GeV  &HE onset & HE emission &            &                &    \\  
\hline
\hline
\multicolumn{9}{|c|}{weak LAT GRBs}\\
\hline
080825C\cite{grb080825c} & long & $\sim$ 10 & 0 & $\circ$ & $\circ$ & x & $\sim$  572 MeV & x\\
081024B\cite{grb081024b} & short & $\sim$ 10 & 2 & $\circ$ & $\circ$ & x & $\sim$ 3.1 GeV & x \\
090217\cite{grb090217}  & long  & $\sim$ 10 & 0 & x & x & x & $\sim$ 866 MeV & x \\
\hline
\hline
\multicolumn{9}{|c|}{bright LAT GRBs}\\
\hline
080916C\cite{grb080916c} & long & $>$ 100 & $>$ 10 & $\circ$ & $\circ$ & $\circ$(weak) & 13.2 GeV & 4.35 \\
090510\cite{grb090510}\cite{grb090510physics}  & short & $>$ 150 & $>$ 20 & $\circ$ & $\circ$ & $\circ$      & 31 GeV   & 0.903 \\
090902B\cite{grb090902b} & long  & $>$ 200 & $>$ 30 & $\circ$ & $\circ$ & $\circ$      & 33 GeV   & 1.822 \\
\hline
\end{tabular}
\label{grbcomp}
\end{center}
\end{table*}

\bigskip 

\end{document}